# A higher-dimensional model of the nucleon-nucleon central potential

Eric R. Hedin

**Dept. of Physics & Astronomy**

**Ball State University, Muncie, IN 47306, USA**

erhedin@bsu.edu

**Abstract.** Based on a theory of extra dimensional confinement of quantum particles [E. R. Hedin, Physics Essays **25**, 2 (2012)], a simple model of a nucleon-nucleon (NN) central potential is derived which quantitatively reproduces the radial profile of other models, without adjusting any free parameters. It is postulated that a higher-dimensional simple harmonic oscillator confining potential localizes particles into 3-d space, but allows for an evanescent penetration of the particles into two higher spatial dimensions. Producing an effect identical with the relativistic quantum phenomenon of *zitterbewegung*, the higher-dimensional oscillations of amplitude $\hbar/mc$ can be alternatively viewed as a localized curvature of 3-d space back and forth into the higher dimensions. The overall spatial curvature is proportional to the particle's extra-dimensional ground state wave function in the higher-dimensional harmonic confining potential well. Minimizing the overlapping curvature (proportional to the energy) of two particles in proximity to each other, subject to the constraint that for the two particles to occupy the same spatial location one of them must be excited into the 1$^{st}$ excited state of the harmonic potential well, gives the desired NN potential. Specifying only the nucleon masses, the resulting potential well and repulsive core reproduces the radial profile of several published NN central potential models. In addition, the predicted height of the repulsive core, when used to estimate the maximum neutron star mass, matches well with the best estimates from relativistic theory incorporating standard nuclear matter equations of state. Nucleon spin, Coulomb interactions, and internal nucleon structure are not considered in the theory as presented in this article.

**Higher dimensional nucleon-nucleon potential**

**1. Introduction and background**

The concept of a potential energy function mediating the nucleon-nucleon (NN) interaction has played a significant role in the history of nuclear physics. Although the NN interaction can be precisely expressed in a quantitative fashion, the physical mechanism of the interaction is not yet understood. Theoretical models for the nuclear force interaction have been historically based upon pion-exchange, beginning with the Yukawa theory [1]. A few decades later, a model based upon the exchange of one heavy meson, or boson (the OBE model) was developed [2]. Correlated with this, a 2-pion exchange contribution has been considered significant for modeling the medium-range part (1 fm < r < 2 fm) of the NN potential, leading to the Paris potential [3] and the Bonn model [4]. The one-pion exchange is still understood as dominating the long-range part (r > 2 fm). The short-range part of the potential (r < 1 fm) has been thought to stem from a complex set of phenomena providing a repulsive core, and has typically been described phenomenologically [3]. Seeking to discover a unified model for the potential which covers all three ranges has proved difficult. Renewed efforts to explain the nuclear potential arose with the advent of quantum chromodynamics (QCD) and quark models known as chiral perturbation theory [5, 6]. Models of this sort initially succeeded in qualitatively reproducing some basic properties of the nuclear force and with higher-order fitting processes have approached the quantitative precision of some phenomenological potentials [7]. Lattice QCD simulations of the effective central part of the NN potential have reproduced a central repulsive core surrounded by an attractive well which also shows consistency with the long-range region described by the one-pion exchange model [8]. The Oxford potential stems from the Non-Relativistic Quark Model, which consists of a combination of virtual particle exchange potentials. It has been shown to be able to reproduce NN scattering phase shifts and properties of the deuteron [9]. A complete microscopic NN potential based on QCD is still a future goal, although a microscopic NN interaction derived from the relativistic mean field theory Lagrangian has given results comparable with phenomenological NN interactions [10]. This potential can also be used to calculate a number of nuclear observables. Recently, a relativistic optical model potential for nucleon-nucleus scattering, which is isospin-dependent has been investigated [11]. Good agreement with experimental scattering data for the case of n, p + $^{27}$Al was obtained. A full parameterization of this potential has, however, not yet been realized.

Although these and other models of the NN potential have attained a refinement which enables their use in describing experimental nuclear data, a first-principles theoretical model which does not rely on the adjustment of free parameters to fit the model to the data has remained elusive. Based on a theoretical framework in which elementary particles have a component of their wave function extending

**Higher dimensional nucleon-nucleon potential**

into higher spatial dimensions, the radial dependence of a NN central potential is herein derived which quantitatively reproduces the standard features of other models in all three spatial regions, without adjusting any free parameters. The higher-dimensional theory upon which this model is based has recently been published elsewhere [12], although a brief summary of the salient points will be reviewed in the following section in order to provide the necessary background for understanding the derivation of the NN potential.

## 2. Higher-dimensional confinement

In Ref. [12] it is proposed that a higher-dimensional (4th and 5th spatial dimensions) simple harmonic oscillator confining potential localizes particles into 3-d space (characterizing the "brane tension" which confines Standard Model particles to the sub-manifold). Quantum effects allow a non-zero probability for a particle's evanescent existence in the higher dimensions. The extent of the higher-dimensions is not constrained in this model, only the confinement scale of particles. In this model, the harmonic potential is used to confine particles and electromagnetic fields to an extremely restricted extra-dimensional region beyond the submanifold. Beyond the extent of this potential confinement, this model does not require a larger size to the extra dimensions. However, the extent of the size of the extra dimension need not be larger than that which is well below experimental constraints. The higher-dimensional component of the particle's wave function is determined by the harmonic-oscillator potential which confines the particle to 3-d space. The ground state wave function of the particle in the higher-dimensional potential well can be expressed in the following form, which also incorporates consistency with the Heisenberg Uncertainty Principle: $\Psi_{0,0}(\gamma,\eta) = Ae^{-(\gamma^2+\eta^2)/2\gamma_0^2}$, where $\gamma$ and $\eta$ are coordinates extending into the 4th and 5th spatial dimensions, $\gamma_0 = \hbar/mc$ is the classical turning point distance, and the normalization, $A = 1/(\gamma_0 \sqrt{\pi})$. The classical angular frequency for this harmonic potential is $\omega_o = mc^2/\hbar$. The particle's ground state energy in the extra dimensional potential well is equal in magnitude to its rest mass energy, $mc^2$. The energy level of the first excited state of the particle in the extra dimensional potential well is $E_1 = 2mc^2$, since for this type of confining potential, the energy levels increase with quantum number, n, as $E_n = \hbar\omega_0(n+1)$, where $n = n_\gamma + n_\eta$; $n_\gamma, n_\eta = 0,1,2,...$, and $\hbar\omega_o = mc^2$. For a particle to jump into the 1st excited state, it must "absorb" an energy equal to its rest mass energy. The probability distribution of the particle in the 1st excited state has a zero at the origin, meaning that the probability of finding the excited particle in 3-d space (at γ=η=0.0) drops transiently to zero.

In Ref. [12], it is proposed that the oscillatory motion of particles in the higher-dimensional potential well can be alternatively viewed as a localized bending of 3-d space back and forth into the

**Higher dimensional nucleon-nucleon potential**

higher dimensions. The nominal radius of spatial curvature localized on a given particle is given by $\gamma_0 = \hbar/mc$: a particle of small mass only shallowly "indents" 3-d space; a heavier particle more sharply bends space. The overall curvature is proportional to the particle's extra-dimensional wave function. Another theoretical treatment which produces a similar concept is the non-relativistic limit of the Dirac equation, in which a term known as the "Darwin term" appears [13]. This term implies that in relativistic theory the electron's position is diffused over a volume of order $(\hbar/mc)^3$, as a consequence of "*zitterbewegung*," which correlates with the oscillatory motion of a particle in the higher-dimensional potential well, having the same magnitude and oscillation frequency.

## 3. Overlapping curvature functions

So far, the discussion has focused on the quantum mechanical treatment of a single particle in its extra-dimensional harmonic potential well. For the case of two free particles brought into near proximity (without assuming any traditional NN interaction), it is postulated that the extra-dimensional harmonic potentials wells of these two particles and their localized spatial curvatures overlap in such a way as to minimize the overall energy of the configuration. The ground-state wave functions of the two particles (which also represent the resulting spatial curvature due to their higher-dimensional oscillations) are formulated so that their Gaussian nature is preserved when they are separated and so that they begin to smoothly merge when brought into proximity. The spatial curvature resulting from the oscillation of the particle in and out of the higher dimensions, as constrained by the extra-dimensional harmonic potential, serves to couple higher-dimensional coordinates with the 3-d coordinates at the position of the particle. The wave functions of each particle are expressed in terms of the two higher-dimensional coordinates $\gamma$ and $\eta$, but the effect of their proximity is also expressible in terms of ordinary spatial coordinates $x$ and $y$, due to the localized curvature of 3-d space resulting from their higher-dimensional oscillations, as noted above. For a given phase of the oscillation of the particle into the higher dimension, the *x-y* plane is taken as the spatial plane perpendicular to the higher-dimensional coordinate along which the oscillation occurs. The spatial separation of the particles will be designated as occurring along the *x*-axis, with one particle arbitrarily positioned at *x*=0, while the position of the other particle is at *x=b*. A 3-d plot of the overlapping curvature functions resulting from the higher-dimensional oscillation of the two particles in proximity to each other is shown in figure 1. For the purposes of the subsequent mathematical analysis, in which the combined energy of the particles is minimized, their curvature functions are expressed in 4 parts related to their spatial separation as follows:

**Higher dimensional nucleon-nucleon potential**

$$\Psi_I(x, y) = Ae^{-x^2/2\gamma_0^2}e^{-y^2/2\gamma_0^2}, \qquad (x \leq 0)$$

$$\Psi_{II}(x, y) = A_1\left(Ae^{-x^2/2\gamma_0^2} + Be^{x^2/2\gamma_0^2}\right)e^{-y^2/2\gamma_0^2}, \qquad (0 < x \leq b/2)$$

$$\Psi_{III}(x, y) = A_1\left(Ae^{-(x-b)^2/2\gamma_0^2} + Be^{(x-b)^2/2\gamma_0^2}\right)e^{-y^2/2\gamma_0^2}, \qquad (b/2 < x < b)$$

$$\Psi_{IV}(x, y) = Ae^{-(x-b)^2/2\gamma_0^2}e^{-y^2/2\gamma_0^2}, \qquad (x \geq b). \qquad (1)$$

$\Psi_I$ and $\Psi_{II}$ comprise the curvature function of the particle located at $x=0$, and $\Psi_{III}$ and $\Psi_{IV}$ comprise the curvature function of the particle located at $x=b$. These functions represent the localized curvature of the particles into the higher dimensions. By having the same $\gamma_0$ factor in each exponent, the two particles are assumed to have the same mass. The constants $B$ and $A_1$ are determined from boundary conditions, and $A = 1/(\gamma_0 \sqrt{\pi})$, as given above. One of the boundary conditions is $\Psi_{II}(b/2, y) = \Psi_{III}(b/2, y)$, which is automatically satisfied by the form of the functions in (1). We also require continuity of the first derivatives of $\Psi_{II}$ and $\Psi_{III}$ with respect to $x$ at $x = b/2$, and for equal-mass particles, symmetry demands that these first derivatives must vanish at $x = b/2$. This condition is satisfied if $B = A\exp(-b^2/4\gamma_0^2)$. The constant $A_1$ is found by requiring $\Psi_{II}(0,0) = \Psi_{III}(b,0) = A$, which yields $A_1 = 1/(1+\exp(-b^2/4\gamma_0^2))$.

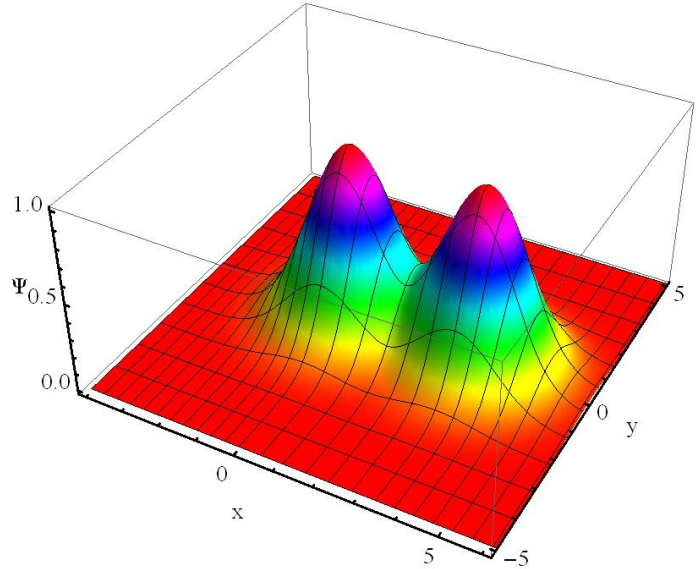

Figure 1 (color online). A plot of the combined curvature functions of two particles at a separation of $r_{sep} = 3.411\gamma_0$. The spatial axes are in units of $\gamma_0$.

**Higher dimensional nucleon-nucleon potential**

## 4. Nucleon-nucleon potential

The next step is to calculate the energy of the two particles as a function of their overlapping curvature and to determine if it decreases as the particles are brought together. Considered from the point of view of a localized region of curvature around each particle, as the particles approach one another the degree of curvature along the line connecting them decreases. The degree of curvature of a function is proportional to its second derivative with respect to the relevant spatial coordinates [14]. The energy of the quantum state of a particle is proportional to the degree of curvature of the particle's wave function. The energy related to the net spatial curvature of the two particles can be calculated from the second derivative of the curvature functions, as shown in the following expression:

$$E_c = \frac{-\hbar^2}{2m}\left[\int_{-\infty}^{0}\Psi_I^*\nabla^2\Psi_I dxdy + \int_{0}^{b/2}\Psi_{II}^*\nabla^2\Psi_{II} dxdy + \int_{b/2}^{b}\Psi_{III}^*\nabla^2\Psi_{III} dxdy + \int_{b}^{\infty}\Psi_{IV}^*\nabla^2\Psi_{IV} dxdy\right], \quad (2)$$

where the curvature functions of the two particles in each region are specified in (1). The limits for the y-integration are from $-\infty$ to $+\infty$. For a single isolated particle, the value of the energy as calculated in (2) is $mc^2/2$, so for two well-separated particles ($b \gg \gamma_0$), the energy as calculated in (2) is $E_c = mc^2$. On the other hand, if the two particles approach each other so that $b=0$, and their wave functions completely overlap, $E_c = mc^2/2$. Thus, the maximum decrease in energy as two particles approach each other from infinity is $\Delta E_c = -mc^2/2$. In counterbalance to this, however, it is postulated that the oscillation of the particle in and out of higher-dimensional space prohibits two particles from occupying the same location, unless one of them is excited to the first excited state of the higher-dimensional harmonic potential well (an energy jump of $mc^2$). Note that this is not the same as an appeal to the Pauli exclusion principle, which, with regards to the origin of the repulsive nuclear core, is argued to not apply to nucleons due to the properties of quantum chromodynamics related to the spin, flavor and color of the six quarks involved in the NN force [15]. The overlap of the curvature functions can be expressed as $S_{12} = \int_{-\infty}^{\infty}\Psi_I \Psi_{IV} dxdy$, where $\Psi_I$ and $\Psi_{IV}$ are given in (1), with the coordinate ranges extended over all space. The overlap integral, $S_{12} = 1.0$ for complete overlap, and $S_{12} \to 0$ for particles infinitely separated. The net change in energy (normalized to zero at infinite separation) as the particles are brought into proximity is $\Delta E_{Tot} = E_c + S_{12}mc^2 - mc^2$. Both $E_c$ and $S_{12}$ are functions of the particles' separation (the parameter, $b$, in (1)). A plot of the net change in energy, $\Delta E_{Tot}$, as a function of the separation distance is given in figure 2. The minimum of the net change in energy

**Higher dimensional nucleon-nucleon potential**

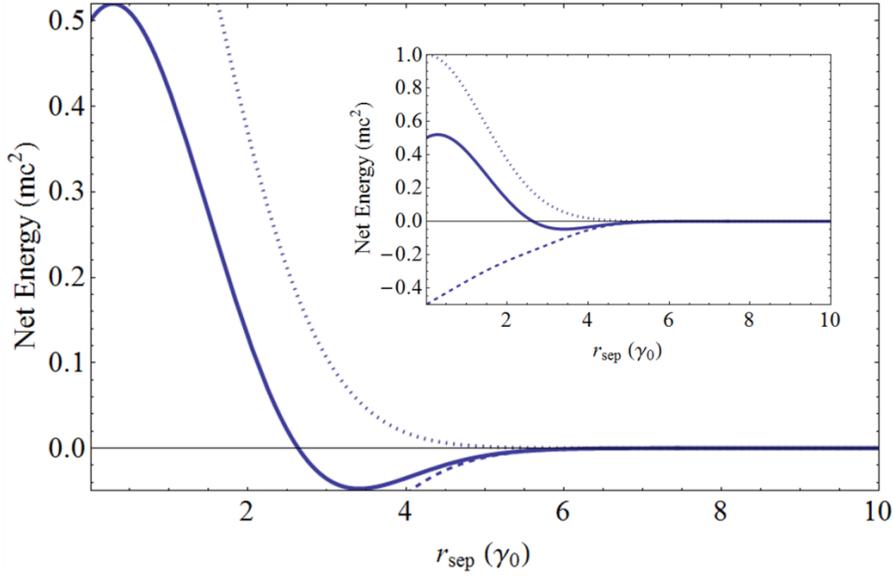

Figure 2  The net change in energy (solid curve) of two equal-mass particles as a function of their separation distance, $r_{sep}$, measured in units of $\gamma_0 = \hbar/mc$. The dotted curve is the overlap energy, $S_{12}mc^2$, and the dashed curve is the normalized curvature energy, $E_c - mc^2$. The inset shows these curves on a larger scale.

occurs at $r_{min} = 3.411\gamma_0$, which is also the value chosen for the separation of the particles depicted in figure 1. The net energy change at $r_{min}$ is $\Delta E_{Tot}(r_{min}) = -0.0473mc^2$, which for a particle mass of 938.9 MeV/$c^2$ (average of the proton and neutron masses) gives the values $\Delta E_{Tot}(r_{min}) = -44.4 MeV$, and $r_{min} = 0.72 fm$. The height of the repulsive core, which increases to about $0.5mc^2$ for $r_{sep} < 0.5 fm$, is about 470 MeV. (For m=939 MeV, the length parameter $\gamma_0 = 0.21 fm$).

## 5. Comparison with standard NN potentials

The plot of $\Delta E_{Tot}$ (solid curve in figure 2) comprises the NN potential as derived directly from the theory of the higher-dimensional quantum confinement of particles in harmonic potential wells. No fitting process of any free parameters in order to facilitate matching to experimental data was incorporated, as is done with phenomenological NN potentials. Comparison with a few other nuclear potential models shows that the higher-dimensional NN potential matches well with the characteristics of the effective

**Higher dimensional nucleon-nucleon potential**

central potentials of those models. It should be mentioned that the higher-dimensional model as presented here does not incorporate any spin-dependence of the nucleons or any Coulomb energy modification. The Paris NN complete singlet central potential as published in [3] shows a minimum energy of $E_{min} = -50 MeV$ at $r_{min} \approx 0.85 fm$. This potential becomes repulsive at $r_{sep} \approx 0.7 fm$. The M3Y effective NN potential (a successful phenomenological NN interaction) as discussed in [10] shows a minimum energy of $E_{min} = -43 MeV$ at $r_{min} \approx 0.78 fm$. This potential becomes repulsive at $r_{sep} \approx 0.6 fm$. Ishii, *et al*, calculate with lattice QCD in Ref. [8] an effective central NN potential with $E_{min} = -30 MeV$ at $r_{min} \approx 0.7 fm$. This potential becomes repulsive at $r_{sep} \approx 0.5 fm$, which matches closely the value shown in figure 2. The Argonne $v_{18}$ central potential [16] has $E_{min} = -60 MeV$ at $r_{min} \approx 0.8 fm$; it becomes repulsive at $r_{sep} \approx 0.7 fm$. Overall, the essential characteristics of the higher-dimensional NN potential lie within the parameter ranges of other sophisticated phenomenological and theoretical models which incorporate extensive free-parameter fitting to experimental nuclear scattering data.

## 6. Neutron star maximum mass

The short-range repulsive core of the NN potential has relevance to neutron stars, in particular to estimates of their maximum mass limit. An accurate physical description of neutron stars requires the use of a General Relativistic approach. The Tolman-Oppenheimer-Volkoff equation [17] incorporates relativistic treatments of the stellar energy density and gravitational binding energy into an equation expressing the equilibrium pressure balance within the star. Although this equation can be solved numerically, the solutions depend upon the equation of state assumed for the compressed nuclear matter. As a consequence, estimates of the maximum stable mass of a neutron star span a range from about 1.7-2.4 solar masses [18], with a broadly applicable argument employing the general theory of relativity indicating a maximum mass of about 3 solar masses [19]. A simpler approach often taken combines quantum statistics and Newtonian gravitational potential in a procedure commonly used to approximate the equilibrium radius of a neutron star [20]. In the treatment provided here, we extend this approach by utilizing the maximum potential of the repulsive core of the higher-dimensional NN potential to obtain an estimate of the neutron star maximum mass. The neutrons, being fermions, obey Fermi-Dirac statistics. The Fermi energy of *N* neutrons within a spherical stellar volume, *V* is

$$E_F = \frac{\hbar^2}{2m_n}\left(\frac{3\pi^2 N}{V}\right)^{2/3}. \quad (3)$$

**Higher dimensional nucleon-nucleon potential**

The average neutron energy is $3E_F/5$, and thus the total neutron energy is, $E_n = 3NE_F/5$. Adding to this the gravitational potential energy, $U_{grav} = -3GM^2/5R$, we obtain the total energy of the star (neglecting thermal energy, and radiative energy) as $E_{Tot} = E_n + U_{grav}$. Minimizing this expression with respect to radius yields an expression for the neutron equilibrium radius as a function of its mass, $(M_{ns} = Nm_n)$,

$$R_{ns} = \left(\frac{3}{2}\right)^{4/3} \pi^{2/3} \frac{\hbar^2}{Gm_n^3} N^{-1/3}. \tag{4}$$

Inserting this expression for $R_{ns}$ into $U_{grav} = -3GM^2/5R$, and substituting $N = 1.2 \times 10^{57}(M_{ns}/M_{Sun})$, the following numerical expression for the gravitational potential energy of the neutron star is obtained:

$$U_{grav} = 8.13 \times 10^{58}(M_{ns}/M_{Sun})^{7/3} \, MeV. \tag{5}$$

The gravitational potential energy per nucleon is $U_{grav}/N = 67.8(M_{ns}/M_{Sun})^{4/3} \, MeV$. The final step is to equate the gravitational potential energy per *pair* of nucleons to the maximum of the repulsive core of the NN potential in order to obtain the maximum mass of the neutron star:

$$2(U_{grav}/N) = 136(M_{ns}/M_{Sun})^{4/3} \, MeV = 0.5m_n c^2 = 470 \, MeV. \tag{6}$$

Solving (6) for $M_{ns}$ yields $M_{ns} = 2.5M_{Sun}$. The value for the mass in the expression for the gravitational potential energy is the mass of the neutron star if its constituents were physically separated to infinity. The effective mass of the star itself (which would be the measured mass of a star) is diminished from this value by the mass-equivalent of the gravitational binding energy, $U_{grav}$ [21]. Using (5) and inserting the value $M_{ns} = 2.5M_{Sun}$ yields $U_{grav} = 6.9 \times 10^{59} \, MeV$. The effective mass is therefore, $M_{ns}^{eff} = M_{ns} - U_{grav}/c^2$. With the accepted value of $M_{Sun} = 1.99 \times 10^{30} \, kg$, we have $U_{grav}/c^2 = 0.6M_{Sun}$, from which we find the effective neutron star maximum mass, $M_{ns}^{eff} = 2.5M_{Sun} - 0.6M_{Sun} = 1.9M_{Sun}$. This estimate of the maximum neutron star mass, although derived via non-relativistic considerations, demonstrates that the higher-dimensional NN potential yields

**Higher dimensional nucleon-nucleon potential**

a value which is consistent with current neutron star theory. For example the average neutron star maximum mass derived from six different realistic models of nuclear forces, as reported in Ref. [18] is $2.0 M_{Sun}$, which is consistent with nearly all known neutron star mass estimates.

## 7. Summary

In conclusion, employing a theory in which a higher-dimensional (4$^{th}$ and 5$^{th}$ spatial dimensions) simple harmonic oscillator confining potential localizes particles into 3-d space, an analytical expression for a simple model of the NN central potential is derived. The potential arises from minimizing the energy associated with the curvature of the overlapping higher-dimensional wave functions of the proximate nucleons, offset by a positive overlap energy resulting from a prohibition against two particles occupying the same location, unless one of them is excited to the first excited state of the higher-dimensional harmonic potential well (an energy jump of $mc^2$). The net energy change as a function of separation distance between the nucleons shows agreement with the radial profile of the NN central potential in several published models. In addition, the maximum energy of the repulsive potential core is obtained, which allows a simple estimate of the maximum neutron star mass. The value obtained is consistent with more sophisticated models employing general relativity and standard nuclear potential models. The consistency of the predictions of this model with the radial profiles of standard NN potential models lends validation to the higher dimensional theory upon which this model is based.

**Higher dimensional nucleon-nucleon potential**